 \documentclass[final,5p,times,twocolumn,authoryear]{elsarticle}

\usepackage{amssymb}
\usepackage{lipsum}
\usepackage{amsmath}
\journal{Applied Thermal Engineering}

\begin{document}

\begin{frontmatter}

%% Title, authors and addresses

%% use the tnoteref command within \title for footnotes;
%% use the tnotetext command for theassociated footnote;
%% use the fnref command within \author or \affiliation for footnotes;
%% use the fntext command for theassociated footnote;
%% use the corref command within \author for corresponding author footnotes;
%% use the cortext command for theassociated footnote;
%% use the ead command for the email address,
%% and the form \ead[url] for the home page:
%% \title{Title\tnoteref{label1}}
%% \tnotetext[label1]{}
%% \author{Name\corref{cor1}\fnref{label2}}
%% \ead{email address}
%% \ead[url]{home page}
%% \fntext[label2]{}
%% \cortext[cor1]{}
%% \affiliation{organization={},
%%            addressline={}, 
%%            city={},
%%            postcode={}, 
%%            state={},
%%            country={}}
%% \fntext[label3]{}

\title{Scientific Machine Learning-assisted Model Discovery from Telemetry Data}

%% use optional labels to link authors explicitly to addresses:
%% \author[label1,label2]{}
%% \affiliation[label1]{organization={},
%%             addressline={},
%%             city={},
%%             postcode={},
%%             state={},
%%             country={}}
%%
%% \affiliation[label2]{organization={},
%%             addressline={},
%%             city={},
%%             postcode={},
%%             state={},
%%             country={}}

\author[1]{Sebastian Micluta-Campeanu}
\author[1]{Avinash Subramanian}
\author[1]{Anas Abdelrehim}
\author[1]{Ranjan Anantharaman}
\author[2]{Rohit Dhumane}
\author[1]{Brad Carman}
\author[1]{Chris Rackauckas}
\affiliation[1]{JuliaHub, Inc.}
\affiliation[2]{Trane Technologies, Inc.}

\begin{abstract}
%% Text of abstract
Calibration of dynamic models to data is an important step in building building digital twins of HVAC equipment, thermal loads and control systems. Sometimes, when a model fails to calibrate to data, a possible cause is that the model has made too many simplifying assumptions and is missing physics. In this paper we propose a semi-automated approach, called Dyad Model Discovery, that can augment the physical equations of the model with symbolic expressions discovered from the data. We demonstrate this method on a digital twin of a transportation refrigeration unit to improve its predictive performance, trained using telemetry data. An engineer-in-the-loop workflow is proposed, which provides suggestions to the user which can then be accepted or rejected. This is the first AI-assisted engineering design workflow to our knowledge.
\end{abstract}

%%Graphical abstract
%\begin{graphicalabstract}
%\includegraphics{grabs}
%\end{graphicalabstract}

%%Research highlights
%\begin{highlights}
%\item Research highlight 1
%\item Research highlight 2
%\end{highlights}

\begin{keyword}
%% keywords here, in the form: keyword \sep keyword, up to a maximum of 6 keywords
model discovery\sep universal differential equations\sep calibration \sep transportation refrigeration

%% PACS codes here, in the form: \PACS code \sep code

%% MSC codes here, in the form: \MSC code \sep code
%% or \MSC[2008] code \sep code (2000 is the default)

\end{keyword}

\end{frontmatter}

%\tableofcontents

%% \linenumbers

%% main text
\section{Introduction}

The calibration of dynamical systems can be a challenging task, especially in the heating ventilation, cooling and refrigeration (HVAC\&R) domain \cite{chakrabarty2021scalable}. A number of calibration approaches may need to explored \cite{rackauckas2022composing} to ensure the convergence of the resulting high-dimensional, non-linear optimization problem. However, the lack of convergence may be due to tangential reasons such as the lack of parameter identifiability \cite{browning2022efficient} or due an insufficiently detailed model that may not capture all of the latent dynamics observed in the data. In this paper, we propose a semi-automated method that takes in an existing model, a dataset and proposes modifications to the model's underlying physical equations. We refer to this process henceforth as model discovery. 

Model discovery is distinct from system identification approaches such as SinDy \cite{fasel2021sindy} or classical linear system identification methods \cite{carlson2021controlsystems}, both of which are purely data-driven. In contrast, our approach is based upon Universal Differential equations proposed in \cite{rackauckas2020universal}, which augments dynamical systems with neural networks, and subsequently calibrates the augmented system to data. Our approach builds upon this, but adds important new features such as sensitivity analysis and symbolic regression, both of which provide unique insight into the physical models developed by the user.

\begin{figure}
    \centering
    \includegraphics[width=0.9\linewidth]{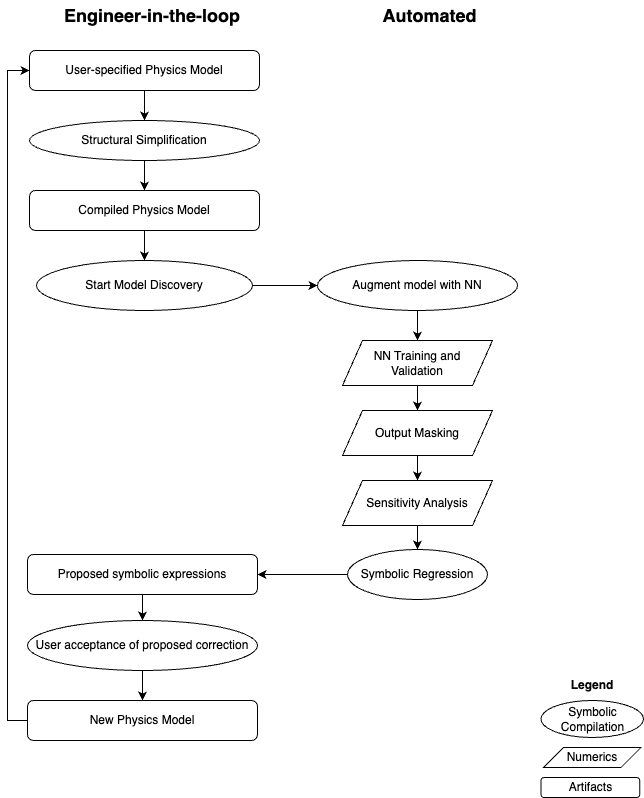}
    \caption{\textbf{Dyad Model Discovery}. The user-facing engineer-in-the-loop steps are on the left, while the automated procedures are found on the right. Finally, the engineer has a choice whether to accept or reject the proposed changes from the system.}
    \label{fig:ude_pipeline}
\end{figure}

In this paper, we demonstrate the usage of this proposed technique on a model of a cargo box, a fundamental modeling component of a transportation refrigeration unit (TRU) \cite{dhumane2023modeling, dhumane2024generic}. Refrigerated transport is a crucial component of the modern economy that ensures the delivery of food and medicine. With increased focus on energy efficiency and decarbonization, there is a need for improved prediction of transport refrigeration unit (TRU) performance for fleet planning and route optimization. The TRU contains a cargo box which is simulated using 1D thermal-fluid models with lumped air control volumes and several parameters like capacitance which are calibrated from test data. The problem is more challenging for multi-temperature cargo where there are several unknown parameters for calibration. To make the problem more challenging, there is significant deviation in the parameters across units. An automatic model calibration or model discovery algorithm can help improve performance prediction of individual units. 

\begin{figure*}
    \centering
    \includegraphics[width=0.5\linewidth]{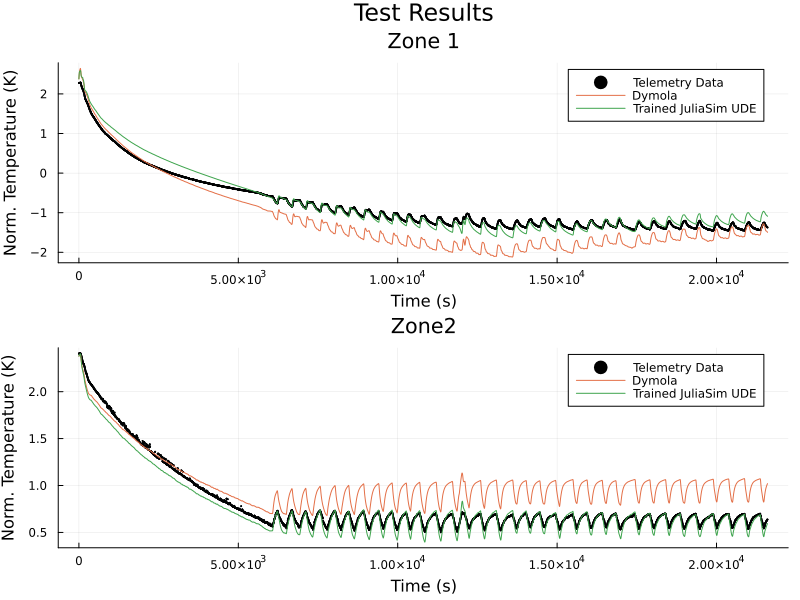}
    \caption{Prediction by a trained Dyad UDE on an unseen dataset, yielding a 3\% improvement in the loss function.}
    \label{fig:results}
\end{figure*}

A transportation refrigeration unit (TRU) consists of a refrigeration system which we shall refer to as the plant, a load which we will refer to as the cargo box, and a control system that ensures that the refrigeration system maintains the load at a particular set point. A detailed description of the physics of the plant, cargo box and control system are given in \cite{dhumane2024generic}. In this paper we shall focus on the cargo box itself, which is the main load of the TRU.

The cargo box is sometimes divided into multiple zones, where each zone is given a different set point, so as to facilitate the transportation of multiple types of cargo during the same trip \cite{dhumane2023modeling}. A multi-zone cargo box consists of multiple cargo box models separated by a bulkhead, through which heat is conducted between the boxes. In this paper, we use a scientific machine learning (SciML) technique to improve the dynamic temperature prediction in a two-zone box model. In particular, we shall use Universal Differential Equations (UDEs) \cite{rackauckas2020universal}, a SciML technique which corrects the physics of a model in order to improve its predictive power. The method is leveraged through the modeling and simulation product Dyad (formerly called JuliaSim) \cite{rackauckas2022composing}, and provides an engineer-guided workflow for fast convergence. 

The rest of this paper is organized as follows: first, we provide background on the UDE method, and describe its implementation and workflow in Dyad, including the different stages of training. Finally, we discuss the implications of the results for future work.

\section{Methods and Tools}
Dyad Model Discovery consists of four numerical methods performed in sequence. In this section, we describe each method and the heuristics used.

\subsection{Universal Differential Equations}

Universal Differential Equations is a scientific machine learning technique used to correct the dynamics of a system of ordinary or differential algebraic equation.  Consider a dynamical physics model of the following form, represented by a differential algebraic equation: 

\begin{align}
    \frac{du_d}{dt} &= f(u, x, p, t) \\
    0 &= g(u, x, p, t)
    \label{eq:dyn}
\end{align}
where  $u(0) = u_0$ and $t \in (0, T)$, and the state vector $u = \{u_d, u_a\}$ where $u_d$ denotes the differential variables and $u_a$ denotes the algebraic variables. Consider the output $y$ from this system depicted by
\begin{align}
    \hat{y}(t) &= h(u, x, p, t)
\end{align}
Consider a test dataset that captures the outputs $\{y(t_i): t_i \in (0, T)\}$, and a residual in the output 

\begin{align}
    \label{eq:loss} r = || \hat{y} - y||_2
\end{align}

The technique tries to correct this residual term by introducing a neural network correction term to the time derivative. The inputs to this neural network are the state derivatives $u_d$ and the outputs are the correction terms required. 

As an example, consider the correction term $NN(u_d, x, p, t; \theta)$, where $\theta$ are the weights of the neural network. This  modifies Equation \ref{eq:dyn}, generating a new dynamical system. 

\begin{align}
\label{eq:ude} \frac{du_d}{dt} &= f(u, x, p, t) + NN(u, x, p, t; \theta) \\ 
   0 &= g(u, x, p, t) \\
   \hat{y} &= h(u, x, p, t)
\end{align}

The optimization problem that is then solved is 
\begin{align}
    \min_\theta r = \min_\theta ||y - \hat{y}||_2 
\end{align}

The above system is posed as an unconstrained optimization problem and can be trained by a number of gradient-based and gradient-free methods. Gradient-based methods would require the implementation of an adjoint equation to calculate the gradient of the residual with respect to the neural network parameters $\theta$. Typically this is implemented using reverse-mode automatic differentiation. An implementation of these adjoint methods is provided in the SciML software stack \cite{rackauckas2020universal, DifferentialEquations.jl-2017}.

\subsubsection{Output Masking}
After training this neural network, the desired correction is in essence determined. The user may choose to simply stop training now and use the resulting grey box model. However, the engineer may choose to reduce the scope of the correction, namely through the following steps: 

\begin{enumerate}
    \item Reducing the number of corrections (output dimension of the network) 
    \item Reducing the number of states needed to make this correction (initially, all states are used for training)
    \item Converting the reduced neural network into symbolic regression.
\end{enumerate}

This section explains how the output dimension of the neural network is reduced. The magnitude of the neural network outputs are examined, and a metric may be chosen to determine if the magnitude of the correction is relatively small. An example heuristic is the ratio of the correction magnitude to the maximum value of the time derivative over time: 
\begin{align*}
    \frac{|\Delta d_i^\prime|}{\max_T |d_i^\prime|}
\end{align*}

In this way, the top outputs may be determined. An iterative masking procedure is then applied where the top X outputs are unmasked and the rest of the outputs are masked. The goal of this procedure is to determine the minimum number of corrections to be kept such that the loss function given by Equation \ref{eq:loss}.

After the number of output dimensions are reduced, the network is re-trained with the smaller output dimension, using the same procedure as described in the previous section. 

\subsubsection{Sensitivity Analysis}
The input dimension may now be reduced using sensitivity analysis. A Jacobian matrix from input to output is computed at the optimal parameters, and the relative magnitudes of the columns can be used to pick the top input variables. 

\subsubsection{Symbolic Regression}
Finally, the neural network can be replaced by symbolic expressions using symbolic regression techniques \cite{cranmer2023interpretable}. This conducts a massive distributed search for expressions that map from the input variables to the output correction values. Model parameters are also included as input variables so as to provide generalizability of the resulting expression across model parameters. Finally, the symbolic regression returns the best equations from the search. 

\subsection{Implementation in Dyad}

Dyad is an acausal modeling environment \cite{rackauckas2022composing, ma2021modelingtoolkit} that composes system modeling and machine learning. Universal Differential Equations are provided via a sub-module of Dyad, known as Dyad AI, which provides an approachable and automated pipeline for using UDEs. Dyad AI consists of both universal differential equations, as well as surrogate modeling methods. \cite{abdelrehim2025active}. The Universal Differential Equations formalism provides augments first-principles physics-based models as well as existing knowledge with a machine learning model that captures unknown physics.

\subsubsection{Physics-based Thermal-fluid modeling}
First, a thermal-fluid model is developed for the TRU in Dyad. The TRU is an insulated box that is divided into 3 different temperature zones using 2 partitions. This Dyad model is based on the low to medium fidelity Modelica-based model developed by Dhumane and Greene (\cite{dhumane2023modeling}). The developed model is a systems of Differential Algebraic Equation (DAEs) generated from composable acausal symbolic models. 

The key dynamic observable quantities of interest are the temperatures in each of the 3 zones. The key physical phenomena of interest are: heat transfer from the ambient air through the insulating walls into the air zones, heat transfer from the air zones to the refrigeration system, and mass transfer within the air zones and between zones resulting from air leaks.

Full details of the physics-based model are presented in (\cite{dhumane2023modeling}). In summary: lumped control volume models are used to represent the air zones, distributed models are used to describe the enclosure walls  by discretizing along the wall thickness (with the other two dimensions kept as lumped control volumes). Convective heat transfer inside the air zone is implemented together with conductive heat transfer through the enclosure walls. Several parameters such as box dimensions (height, width, zone lengths) as well as heat and mass transfer correlation parameters influence the model with the latter determined using a calibration analysis. The two zones are separated by means of a bulkhead, which accounts for conductive heat transfer between the zones.

\subsubsection{Structural Simplification}
A physics model in Dyad is fundamentally a system of differential-algebraic equations. This system, as specified by the user, undergoes a series of transformations and symbolic manipulations \cite{ma2021modelingtoolkit} such as index lowering via the Pantiledes algorithm, alias elimination, tearing, dummy derivatives and others. These symbolic manipulations preserve the system exactly and are not approximations. The result is a smaller, symbolically simplified system of equations.

\begin{figure*}
    \centering
    \includegraphics[width=0.8\linewidth]{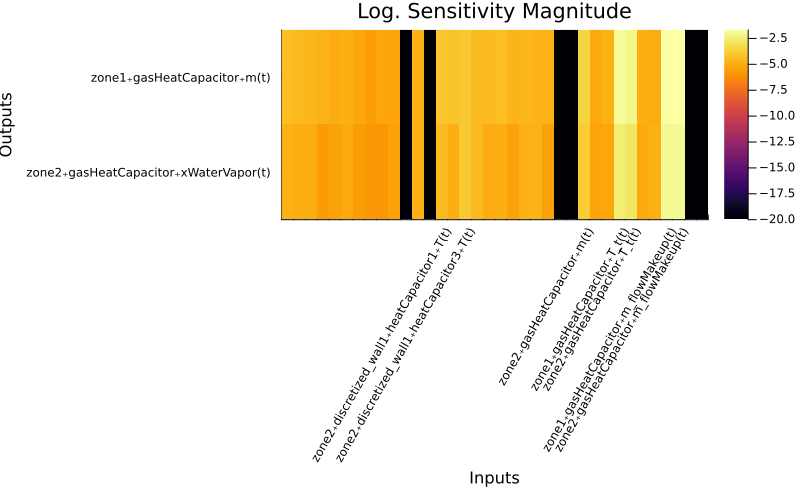}
    \caption{A heatmap of the jacobian matrix computed as part of the sensitivity analysis. The top 7 inputs are marked on the X-axis. Inputs that end in "\_t(t)" represent dummy derivatives generated by the model compilation.}
    \label{fig:sens}
\end{figure*}

\subsubsection{Universal Differential Equations}

After the Dyad model is structurally simplified, the dynamics are now augmented with a neural network. Initally, the input dimension of the neural network is the number of states and the output is the number of differential equations/variables. Before training  an input and output normalization step is undertaken: input normalization prevents the neural network activation functions from saturating and output normalization steps ensure the correction terms are not too large so as to make the augmented system unstable. After the normalization steps, the neural network is trained such that the solution of the hybrid system matches the data well. Additionally, a number of different architectures are trained simultaneously using an experiment tracking system within Dyad. The architecure with the best performance is automatically picked for the user, but the engineer can go into the experiment tracker and pick a different neural network if desired. 

\subsubsection{Masking, Sensitivity Analysis and Symbolic Regression}

Once the optimization problem has converged to a solution, a new hybrid system with the neural network embeddings is produced. Two heuristics are performed to reduce the size of the neural network: the outputs with the highest magnitudes are considered, while masking the low magnitude corrections. This reduces the output dimension of the neural network. The network is then re-trained with the smaller output dimension. 

Next, sensitivity analysis step is performed to determine which variables most significantly contribute to these corrections. This reduces the input dimension of the correction. Finally, a symbolic regression \cite{cranmer2023interpretable} is performed to replace the input-output of the neural network with symbolic expressions. The reduced outputs, reduced input variables, along with hand-chosen parameters are fed to the symbolic regression algorithm, which then produces a minimal expression. The final result is a new model which consists of new symbolic expressions can be appended to the original model, and the choice is provided to the user. The entire pipeline is summarized in Figure \ref{fig:ude_pipeline}.

% In Dyad, UDEs are implemented as a code injection at a particular stage of the Dyad compilation chain. A typical Dyad model, which is set a differential-algebraic equations undergoes a series of compiler transformations and simplifications which turns it into an ordinary differential equation. It is at this stage that the neural network is augmented to the system dynamics, and trained against test data. Once the neural network is trained, a symbolic regression step is performed to turn the data-driven neural network turn into a weighted combination of symbolic terms. This not only adds some interpretability to the resulting system, but also improves its generalizability. 

% In the following subsections we describe briefly describe the various automated stages of the UDE training process, starting with the a Dyad model specified by the user. 

\begin{table*}
\centering
\begin{tabular}{|c|c|c|} 
 \hline
 Variable & Type & Description \\  
 \hline\hline
 xPartition & Parameter & Relative size of two partitions  \\
 initialAirTemperatureZ1 & Parameter & Initial Temperature of Zone 1 \\
 initialAirTemperatureZ2 & Parameter & Initial Temperature of Zone 2 \\
 zone1.gasHeatCapacitor.m\_flowMakeup & Differential Variable & Mass Flow Rate in Zone 1 \\
 zone2.gasHeatCapacitor.m\_flowMakeup & Differential Variable & Mass Flow Rate in Zone 2 \\
 zone2.discretized\_wall1.heatCapacitor1.T & Differential Variable & Temperature in Discrete Element of Wall \\
 zone2.discretized\_wall1.heatCapacitor3.T & Parameter & Temperature in Discrete Element of Wall \\
 zone2.gasHeatCapacitor.m & Algebraic Variable & Mass Within Zone2 \\
 zone1.gasHeatCapacitor.T\_t & Dummy Derivative & D.D of Temperature in Zone 1 \\
 zone2.gasHeatCapacitor.T\_t & Dummy Derivative & D.D of Temperature in Zone 1 \\
 \hline\hline
\end{tabular}
\caption{Inputs variables supplied to the the symbolic regression routine, decided based on the results of the sensitivity analysis and some knowledge of the model.}
\label{tab:dataset}
\end{table*}

\section{Results} 

\subsection{Test Model, Dataset and Structural Simplification}
\begin{figure}
    \centering
    \includegraphics[width=0.5\linewidth]{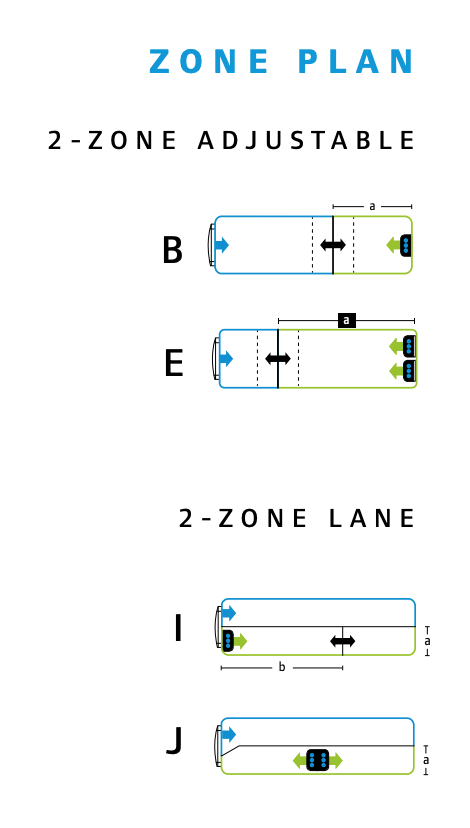}
    \caption{Trailer configurations for a two-zone box model. The Dyad UDE was trained on Configuration B and E, and tested on a proprietary configuration.}
    \label{fig:config}
\end{figure}

A detailed description of a multizone cargo box model is available in \cite{dhumane2023modeling} along with the discretization schemes employed to build the model. Our test model is a two-zone cargo box model with thermal insulation on all sides. Zones are thermally separated by means of a bulkhead. The air within the boxes are modeled as a lumped control volume. Heat transfer occurs between zones due to temperature difference and mass transfer occurs due to temperature difference. The two zone box model also has several configurations corresponding to the relative sizes of each zone. Our training dataset corresponded to the following configurations B and E as shown in Figure  \ref{fig:config}

More specifically, the test cases that were studied are summarized in Table \ref{tab:dataset} and Figure \ref{fig:config}. The dataset consists of three instances of telemetry data corresponding to the temperature in each zone. We use two of the three data instances for training and one for test.  

The Dyad model underwent structural simplification, generating a compiled model with 36 states of which 8 were algebraic variables and 28 were differential variables. 

\begin{table}
\centering
\begin{tabular}{|c|c|c|c|c|} 
 \hline
 ID & \multicolumn{2}{|c|}{Dimensions} & \multicolumn{2}{|c|}{Temperature} \\ [0.7ex] 
 & Zone1 & Zone2 & Zone1 & Zone2\\
 \hline\hline
 1 & 9m & 4.5m & -20C & 0C \\
 2 & 2.5m & 11m & -25C & 4C\\
 3 & 4.5m & 9m & -20C & 2C\\
 \hline\hline
\end{tabular}
\caption{Different vehicle configurations in the training set (\#1 and \#3) and test set (\#2)}
\label{tab:dataset}
\end{table}

\subsection{Universal Differential Equations}

As a starting point, a neural network $\mathcal{N}: u \mapsto \Delta u_d^\prime$ is appended to the system like Equation \ref{eq:ude}. The input dimension of the neural network was the total number of states in the system (36) and the output dimension was the total number of differential equations in the system (28). A number of different hidden layers and architectures are trained using Dyad's experiment tracking feature and the best architecture was chosen. For this model, multi-layer perceptrons with different hidden layer sizes were trained. The final fitting results are shown in Figure \ref{fig:results}.

% In order to be able to parametrize the generated approximation based on the zone lengths, we used the first two cases (ID 1 and 2) for training neural networks and the last one to validate the results.
% For the UDE methodology we first have to augment the structurally simplified system of differential algebraic equations (DAEs) with a neural network on the right hand side, such that the inputs to the neural network are the values of the corresponding variables and the output, or the prediction of the neural network is added to the terms representing the differential variables in the DAE system.

% As such the neural network represents a function that takes in all the non-constant variables in the system of DAEs and outputs values that represent derivative correction terms to be added to the right hand side of the equations for each of the differential variables in the system. To concretely implement neural network augmentation the Dyad software was used.

\subsection{Masking and Sensitivity Analysis}

In order to improve the generalizability of the resulting UDE, Dyad is determining what are the most influential equations in the system and produces neural network output data only for those specific equations. The trained neural network from the previous step was analyzed. Of the 28 outputs, it was found that most of the neural network outputs were actually close to zero, and the outputs with the top 2 magnitudes provided sufficient correction for a very low loss value. 

Thus, a new network with 36 inputs and 2 outputs was constructed and trained using the same procedure as before, resulting in the same predictive performance as the previous network. Finally, a sensitivity analysis is conducted by computing a jacobian matrix of the outputs to the inputs, to determine the input variables that the outputs are most sensitive to. A heatmap of the jacobian matrix is shown in Figure \ref{fig:sens}.

\subsection{Symbolic Regression}

That data is then further processed with symbolic regression \cite{cranmer2023interpretable}, such that we obtain parametrized expressions that represent the UDE correction terms. The symbolic regression requires a set of input variables and parameters. The top 7 inputs as determined by the sensitivity analysis as well as three parameters were provided to the symbolic regression and an expression was generated. Table X shows the inputs provided to the symbolic regression. 

\begin{figure*}
    \centering
    \includegraphics[width=1.0\linewidth]{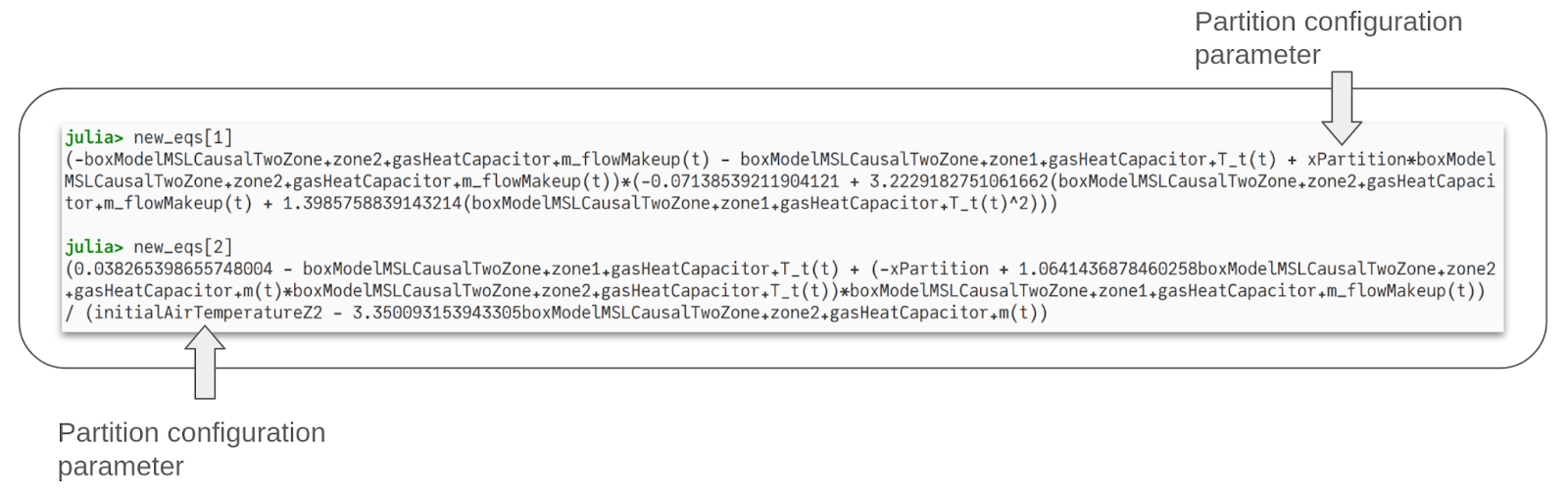}
    \caption{The final symbolic expressions generated by the symbolic regression. Only five input variables and two parameters were used to generate the expressions.}
    \label{fig:symbolic_regression}
\end{figure*}

Fig \ref{fig:symbolic_regression} shows the resulting correction terms that were obtained from the symbolic regression procedure on the first two datatest. As we can observe, the symbolic expressions are parametrized by zone configuration specific parameters, such as the initial air temperature in zone 2 or the length of the partitions. Using symbolic expression for the correction terms enables us to have a more general correction term. In order to test the results, the last test case was used.

\section{Discussion}

The Dyad-UDE workflow makes use of a number of symbolic and numeric passes, and so naturally a model will have to ported to Dyad to take advantage of this method. This can be accomplished either through either the Dyad graphical program or by writing code in the Dyad Modeling Language (JSML), an acausal declarative language embedded in Julia. In this instance, the test model was ported over from the Modelica language to JSML. 

Model updates are not performed without the engineer's approval. The engineer also has the freedom to make choices at every stage of the training process, such as picking a suitable architecture from a list of hyperparameter searches, the number of output variables to mask, and finally, most notably the input variables into the symbolic regression. In this case, we made a choice to include three parameters into the regression, and this choice was informed by knowledge of the model itself. This underscores the need for the engineer to be involved. 

Each training stage of the neural network involved testing over 500 different choices of architecure and hyperparameters through an experiment tracker. Every training run was submitted to the JuliaHub cloud product and linked to the experiment tracker through a job ID. 

After the symbolic regression, we notice that dummy derivatives show up in the final expressions, as they come from the prior sensitivity analysis. Dummy derivatives are a-physical and are simply an artifact of the compilation process. Future work is required to determine how to eliminate them in favor of physical quantities. 

Finally, these corrections are applied to a compiled model (i.e., post structural simplification). Future work is required to determine how to correct specific equations in the user code before compilation. 

\section{Conclusion}
Model Discovery is a powerful tool to autocomplete system models. In this work, model discovery was able to fit the data accurately and extrapolate to unseen conditions. They are a promising tool to improve the accuracy of component based models, and a workflow is presented within Dyad.

\bibliographystyle{elsarticle-harv} 
\bibliography{references}

\end{document}